\documentclass[aps,prd,amsmath,amssymb,preprintnumbers,12pt,twoside, a4paper]{article}

\usepackage[dvipdfmx]{graphicx}
\usepackage{amsmath,amssymb,graphicx,epsfig}
\usepackage[hdivide={25mm,,25mm}, vdivide={25mm,,25mm}, nohead]{geometry}

\usepackage[normalem]{ulem} 
\usepackage{cancel} 

\usepackage{color}

\newcommand{\Tr}{\text{Tr}}

\usepackage{subfigure}
\usepackage{fancyhdr}
\usepackage{enumitem}
\def\simlt{\mathrel{\lower2.5pt\vbox{\lineskip=0pt\baselineskip=0pt
             \hbox{$<$}\hbox{$\sim$}}}}
\def\simgt{\mathrel{\lower2.5pt\vbox{\lineskip=0pt\baselineskip=0pt
             \hbox{$>$}\hbox{$\sim$}}}}

\input{epsf.sty} 
\begin{document}

\thispagestyle{empty}

\begin{center}
{\Large{\textbf{Inflation from Supersymmetry Breaking}}}
\\
\medskip
\vspace{1cm}
\textbf{
I.~Antoniadis$^{\,a,b,}$\footnote{antoniadis@itp.unibe.ch}, 
A.~Chatrabhuti$^{c,}$\footnote{dma3ac2@gmail.com}, 
H.~Isono$^{c,}$\footnote{hiroshi.isono81@gmail.com}, 
R.~Knoops$^{c,}$\footnote{rob.k@chula.ac.th}
}
\bigskip

$^a$ {\small LPTHE, UMR CNRS 7589
Sorbonne Universit\'es, UPMC Paris 6,\\ 4 Place Jussieu, 75005 Paris, France}

$^b$ {\small Albert Einstein Center, Institute for Theoretical Physics,
University of Bern,\\ Sidlerstrasse 5, CH-3012 Bern, Switzerland }

$^c$ {\small Department of Physics, Faculty of Science, Chulalongkorn University,
\\Phayathai Road, Pathumwan, Bangkok 10330, Thailand }

\end{center}

\vspace{1cm}

\begin{abstract}
 We explore the possibility that inflation is driven by supersymmetry breaking with the superpartner of the goldstino (sgoldstino) playing the role of the inflaton. 
 Moreover, we impose an R-symmetry that allows to satisfy easily the slow-roll conditions, avoiding the so-called $\eta$-problem, 
 and leads to two different classes of small field inflation models; they are characterised by an inflationary plateau around the maximum of the scalar potential, 
 where R-symmetry is either restored or spontaneously broken, with the inflaton rolling down to a minimum describing the present phase of our Universe. 
 To avoid the Goldstone boson and remain with a single (real) scalar field (the inflaton), R-symmetry is gauged with the corresponding gauge boson becoming massive. 
 This framework generalises a model studied recently by the present authors, 
 with the inflaton identified by the string dilaton and R-symmetry together with supersymmetry restored at weak coupling, at infinity of the dilaton potential. The presence of the D-term allows a tuning of the vacuum energy at the minimum. 
 The proposed models agree with cosmological observations and predict a tensor-to-scalar ratio of primordial perturbations $10^{-9}\simlt r\simlt 10^{-4}$
 and an inflation scale $10^{10}$ GeV $\simlt H_*\simlt 10^{12}$ GeV. $H_*$ may be lowered up to electroweak energies only at the expense of fine-tuning the scalar potential.
\end{abstract}

\newpage
\section{Introduction}
Inflationary models~\cite{Guth:1980zm}
in supergravity\footnote{For reviews on supersymmetric models of inflation, see for example~\cite{Lyth:1998xn}.}
suffer in general from several problems, such as fine-tuning to satisfy the slow-roll conditions, large field initial conditions that break the validity of the effective field theory, and stabilisation of the (pseudo) scalar companion of the inflaton arising from the fact that bosonic components of superfields are always even. The simplest argument to see the fine tuning of the potential is that a canonically normalised kinetic term of a complex scalar field $X$ corresponds to a quadratic K\"ahler potential $K=X{\bar X}$ that brings one unit contribution to the slow-roll parameter $\eta=V''/V$, arising from the $e^K$ proportionality factor in the expression of the scalar potential $V$. 
This problem can be avoided in models with no-scale structure where cancellations arise naturally due to non-canonical kinetic terms leading to potentials with flat directions (at the classical level). 
However, such models require often trans-Planckian initial conditions that invalidate the effective supergravity description during inflation. 
A concrete example where all these problems appear is the Starobinsky model of inflation~\cite{Starobinsky:1980te}, despite its phenomenological success.

In this work we show that all three problems above are solved when the inflaton is identified with the scalar component of the goldstino superfield\footnote{
See~\cite{Randall:1994fr} for earlier work relating supersymmetry and inflation.}, in the presence of a gauged R-symmetry. Indeed, the superpotential is in that case linear and the big contribution to $\eta$ described above cancels exactly. Since inflation arises in a plateau around the maximum of the scalar potential (hill-top) no large field initial conditions are needed, while the 
pseudo-scalar companion of the inflaton is absorbed into the R-gauge field that becomes massive, leading the inflaton as a single scalar field present in the spectrum. This model provides therefore a minimal realisation of natural small-field inflation in supergravity, compatible with present observations, as we show below. Moreover, it allows the presence of a realistic minimum describing our present Universe with an infinitesimal positive vacuum energy arising due to a cancellation between an F- and D-term contributions to the scalar potential, without affecting the properties of the inflationary plateau, 
along the lines of Ref.~\cite{Antoniadis:2014hfa,Antoniadis:2016aal,Villadoro:2005yq}.

On general grounds, there are two classes of such models depending on whether the maximum corresponds to a point of unbroken (case 1) or broken (case 2) R-symmetry. The latter corresponds actually to a generalisation of the model we studied recently~\cite{Antoniadis:2016aal}, inspired by string theory~\cite{Antoniadis:2014hfa}. It has the same field content but in a different field basis with a chiral multiplet $S\propto \ln X$ playing the role of the string dilaton. Thus, $S$ has a shift symmetry which is actually an R-symmetry gauged by a vector multiplet and the superpotential is a single exponential. The scalar potential has a minimum with a tuneable vacuum energy
and a maximum that can produce inflation when appropriate corrections are included in the K\"ahler potential. 
In these coordinates R-symmetry is restored at infinity, corresponding to the weak coupling limit. As we show below, this model can be generalised to a class of models (case 2) where inflation arises at a plateau where (gauged) R-symmetry is spontaneously broken. 
Small field inflation is again guaranteed consistently with the validity of the effective field theory.
In this work,
we are mostly focused on a new possibility (case 1) where the maximum is around the origin of $X$ where R-symmetry is restored.

The outline of the paper is the following: In Section~\ref{model}, we describe the model and the two cases to study. 
In Section~3, we analyse in detail case 1. We first work out model independent predictions valid under the assumption that inflation ends very rapidly after the inflationary plateau of the scalar potential (section 3.1). We then discuss the possibility for the existence of a minimum nearby the maximum with an infinitesimal tuneable positive vacuum energy and show that additional corrections to the K\"ahler potential are needed (section 3.2). The notion of nearby is defined in the sense that perturbative expansion around the maximum is valid for the K\"ahler potential, but not for the slow-roll parameters. We subsequently present an explicit example and work out the cosmological predictions (section 3.3). Finally, in section 3.4, we discuss the anomaly cancellation associated to the $U(1)_R$ gauge symmetry. In Section~\ref{sec:case_2} we analyse case 2, firstly in general (section 4.1) and then in a particular example. In Section~\ref{sec:no_TeV} we derive a stringent constraint on low energy inflation models in a model-independent way. Section~\ref{sec:no_TeV} can be read independently of the rest of the paper. Our conclusions are presented in Section~6, while Appendix~A contains a proof of an identity used in Section~\ref{sec:no_TeV}.

\section{Symmetric versus non-symmetric point}\label{model}

In this work we are interested in supergravity theories containing a single chiral multiplet transforming under a gauged R-symmetry with a corresponding abelian vector multiplet. 
We assume that the chiral multiplet $\mathcal X$ (with scalar component $X$) transforms as:
\begin{align}
 X \longrightarrow X e^{-iq \omega}. \label{U1}
\end{align}
where $q$ is its charge, and $\omega$ is the gauge parameter.

The K\"ahler potential is therefore a function of $X\bar X$, while the superpotential is constrained to be of the form $X^b$:
\begin{align}
 \mathcal K = \mathcal K(X \bar X), \notag \\
 W = \kappa^{-3} f X^b , \label{Case1:General}
\end{align}
where $X$ is a dimensionless field and $\kappa^{-1}  = m_p = 2.4 \times 10^{15} \text{ TeV}$ is the (reduced) Planck mass.
For $b \neq 0$, the gauge symmetry eq.\,\eqref{U1} becomes a gauged R-symmetry. 
The gauge kinetic function can have a constant contribution as well as a contribution proportional to $\ln X$
\begin{align}
 f(X) = \gamma + \beta \ln X.  \label{case1:gaugekin}
\end{align}
The latter contribution proportional to $\beta$ is not gauge invariant and can be used as a Green-Schwarz counter term to cancel possible anomalies. 
We will show however below, in Section \ref{sec:Anomaly}, that the constant $\beta$ is fixed to be very small by anomaly cancellation conditions and does not change our results.
We will therefore omit this term in our analysis below.

We are interested in the general properties of supergravity theories of inflation that are of the above form.
Before performing our analysis, a distinction should be made concerning the initial point where slow-roll inflation starts. 
The inflaton field (which will turn out to be $\rho$, where $X = \rho e^{i \theta}$) can either have its initial value close to the symmetric point where $X = 0$, 
or at a generic point $X \neq 0$. 
The minimum of the potential, however, is always at a nonzero point $X \neq 0$.
This is because at $X=0$ the negative contribution to the scalar potential vanishes and no cancellation between F-term and D-term is possible.
The supersymmetry breaking scale is therefore related to the cosmological constant as $\kappa^{-2} m_{3/2}^2 \approx \Lambda$.
One could in principle assume that the value of the potential at its minimum is of the supersymmetry breaking scale. 
However, in this case additional corrections are needed to bring down the minimum of the potential to the present value of the cosmological constant, and we therefore do not discuss this possibility.  

In the first case, inflation starts near $X=0$, and the inflaton field will roll towards a minimum of the potential at $X \neq 0$.
On the other hand, in the second case inflation will start at a generic point $X \neq 0$.
In order to make easier contact with previous literature~\cite{Antoniadis:2014hfa,Antoniadis:2016aal}, 
it is convenient to work with another chiral superfield $S$, which is invariant under a shift symmetry
\begin{align}
S \longrightarrow S - i c \alpha \label{shift_symmetry}
\end{align}
by performing a field redefinition
\begin{align}
 X = e^S. \label{field_redef}
\end{align}
In this case the most general K\"ahler potential and superpotential are of the form 
\begin{align}
 \mathcal K &= \mathcal K (S + \bar S), \notag \\
 W &= \kappa^{-3} a e^{bS} .
\end{align}
Note that this field redefinition is not valid at the symmetric point $X=0$ for the first case.

The first case will be discussed in Section \ref{sec:case_1}, and the second case will be discussed in Section \ref{sec:case_2}.


\section{Case 1: Inflation near the symmetric point} \label{sec:case_1}

\subsection{Slow roll parameters} \label{sec:case_1_slowrolls}
In this section we derive the conditions that lead to slow-roll inflation scenarios, where the start of inflation is near a local maximum of the potential at $X=0$.
Since the superpotential has charge 2 under R-symmetry, one has $\langle W \rangle = 0$ as long as R-symmetry is preserved. 
 Therefore, $\langle W \rangle$ can be regarded as the order parameter of R-symmetry breaking.
 On the other hand, the minimum of the potential requires $\langle W \rangle \neq 0$ and broken R-symmetry. 
 It is therefore attractive to assume that at earlier times R-symmetry was a good symmetry, switching off dangerous corrections to the potential.
 As similar approach was followed in~\cite{Schmitz:2016kyr}, where a discrete R-symmetry is assumed.
 Instead, we assume a gauged R-symmetry which is spontaneously broken at the minimum of the potential.

While the superpotential is uniquely fixed in eq.\,(\ref{Case1:General}), the K\"ahler potential is only fixed to be of the form $\mathcal K (X \bar X)$.
We expand the K\"ahler potential as follows
\begin{align}
 \mathcal K(X,\bar X) &= \kappa^{-2} X \bar X + \kappa^{-2} A (X \bar X)^2, \notag \\
  W(X) &= \kappa^{-3}f X^b , \notag \\
 f(X) &= 1, \label{case1_defs}
\end{align}
where $A$ and $f$ are constants. 
The gauge kinetic function is taken to be constant since it will be shown in Section \ref{sec:Anomaly}
that the coefficient $\beta$ in front of the logarithmic term in eq.\,(\ref{case1:gaugekin}) is fixed to be very small by anomaly cancellation conditions.
As far as the scalar potential is concerned, the coefficient $\gamma$ can be absorbed in other parameters of the theory. We therefore take $\gamma = 1$.

The scalar potential is given by\footnote{We follow the conventions of~\cite{Freedman:2012zz}.}
\begin{align}
 \mathcal V  = \mathcal V_F + \mathcal V_D,
\end{align}
where 
\begin{align}
\mathcal V_F = \kappa^{-4} f^2 (X \bar X)^{b-1} e^{X \bar X \left( 1 + A X \bar X \right) }
 \left[-3 X \bar X + \frac{\left( b + X \bar X (1 + 2 A X \bar X) \right)^2}{1 + 4 A X \bar X} \right], 
\end{align}
and
\begin{align}
  \mathcal V_D = \kappa^{-4} \frac{q^2}{2} \left[ b + X \bar X (1 + 2 A X \bar X) \right]^2 .
\end{align}

The superpotential is not gauge invariant under the $U(1)$ gauge symmetry. Instead it transforms as
\begin{align}
 W \rightarrow W e^{-iq b w}\, .\label{Wtransform}
\end{align}
Therefore, the $U(1)$ is a gauged R-symmetry which we will further denote as $U(1)_R$. 
From $W_X k^X_R = - r_R \kappa^2 W$, where $k^X_R = -iq  X$ is the Killing vector for the field $X$ under the R-symmetry,
 $r_R = i \kappa^{-2} \xi_R$ with $\kappa^{-2} \xi_R$ the Fayet-Iliopoulos contribution to the scalar potential, and $W_X$ is short-hand for $\partial W / \partial X$,
we find 
\begin{align}
 r_R &= i \kappa^{-2} q b .
\end{align}
A consequence of the gauged R-symmetry is that the superpotential coupling $b$ enters the D-term contribution of the scalar potential as a constant Fayet-Iliopoulos contribution.\footnote{
For other studies of inflation involving Fayet-Iliopoulos terms see for example~\cite{Binetruy:1996xj}, or~\cite{FI-inflation} for more recent work.
Moreover, our motivations have some overlap with~\cite{Schmitz:2016kyr}, where inflation is also assumed to start near an R-symmetric point at $X=0$.
However, this work uses a discrete R-symmetry which does not lead to Fayet-Iliopoulos terms. 
}

Note that the scalar potential is only a function of the modulus of $X$ and that the potential contains a Fayet-Iliopoulos contribution for $b \neq 0$.
Moreover, its phase will be `eaten' by the $U(1)$ gauge boson upon a field redefinition of the gauge potential similarly to the standard Higgs mechanism.
After performing a change of field variables 
\begin{align}
X = \rho e^{i \theta}, \quad \bar X = \rho e^{-i \theta}, \quad (\rho\ge 0)
\end{align}
the scalar potential is a function of $\rho$,
\begin{align}
 \kappa^4 \mathcal V = f^2 \rho^{2(b-1)} e^{\rho^2 + A \rho^4}  \left( - 3 \rho^2 + \frac{\left( b + \rho^2 + 2 A \rho^4\right)^2}{1 + 4 A \rho^2} \right) + \frac{q^2}{2} \left( b + \rho^2 + 2 A \rho^4 \right)^2 . \label{case1_scalarpotb}  
\end{align}
Since we assume that inflation starts near $\rho=0$, we require that the potential eq.\,(\ref{case1_scalarpotb}) has a local maximum at this point.
It turns out that the potential only allows for a local maximum at $\rho=0$ when $b=1$.
For $b<1$ the potential diverges when $\rho$ goes to zero. For $1 < b < 1.5$ the first derivative of the potential diverges,
while for $b=1.5$, one has $V'(0) = \frac{9}{4}f^2 + \frac{3}{2}q^2 > 0$, and for $b> 1.5$, on has $V''(0) > 0$.
We thus take $b=1$ and the scalar potential reduces to
\begin{align}
 \kappa^4 \mathcal V = 
f^2 e^{\rho^2 + A \rho^4} 
 \left( - 3 \rho^2 + \frac{\left( 1 + \rho^2 + 2 A \rho^4\right)^2}{1 + 4 A \rho^2} \right)
 + \frac{q^2}{2} \left( 1 + \rho^2 + 2 A \rho^4 \right)^2 . \label{case1_scalarpot}  
\end{align}
Note that in this case the the superpotential is linear $W = f X$, describing the sgoldstino (up to an additional low-energy constraint)~\cite{SGoldstino}.
Indeed, modulo a D-term contribution, the inflaton in this model is the superpartner of the goldstino. 
In fact, for $q=0$ the inflaton reduces to the partner of the goldstino as in Minimal Inflation models~\cite{SGoldstino_Inflation}.
The important difference however is that this is a microscopic realisation of the identification of the inflaton with the sgoldstino,
and that the so-called $\eta$-problem is avoided (see discussion below).

The kinetic terms for the scalars can be written as\footnote{The covariant derivative is defined as $\hat \partial_\mu X = \partial_\mu X - A_\mu k^X_R$, where $k^X_R=-iqX$ is the Killing vector for the ${\rm U}(1)$ transformation eq.~\eqref{U1}.}
\begin{align}
  \mathcal L_{\text{kin}} 
 = - g_{X \bar X} \hat \partial_\mu X \hat \partial^\mu X
  = - g_{X \bar X} \left[ \partial_\mu \rho \partial^\mu \rho + \rho^2 \left( \partial_\mu \theta + q A_\mu \right) \left( \partial^\mu \theta + q A^\mu \right) \right] . 
 \label{Case1:Lkin}
 \end{align}
It was already anticipated above that the phase $\theta$ plays the role of the longitudinal component of the gauge field $A_\mu$, which acquires a mass by a Brout-Englert-Higgs mechanism.

We now interpret the field $\rho$ as the inflaton.
It is important to emphasise that, 
in contrast with usual supersymmetric theories of inflation where one necessarily has two scalar degrees of freedom resulting in multifield inflation~\cite{Baumann:2011nk},
our class of models contains only one scalar field $\rho$ as the inflaton.
In order to calculate the slow-roll parameters, one needs to work with the canonically normalised field $\chi$ satisfying
 \begin{align}
 \frac{d\chi}{d\rho} = \sqrt{2 g_{X \bar X} }. \label{case1_norm}
\end{align}
The slow-roll parameters are given in terms of the canonical field $\chi$ by
\begin{align}
\epsilon = \frac{1}{2\kappa^2} \left( \frac{dV/d\chi}{V}\right)^2, \quad
 \eta = \frac{1}{\kappa^2} \frac{d^2V/d\chi^2}{V}. 
 \label{slowroll_pars2}
\end{align}
Since we assume inflation to start near $\rho=0$, we expand
\begin{align}
 \epsilon &=  4 \left( \frac{-4 A  + x^2}{2 + x^2} \right)^2 \rho^2
 + \mathcal O(\rho^4),  \notag \\
 \eta &= 2 \left( \frac{-4 A  +  x^2}{2  + x^2} \right) + \mathcal O(\rho^2) \label{case1_slowroll} ,
\end{align}
where we defined $q = f x$.
Notice that for $\rho \ll 1$ the $\epsilon$ parameter is very small, 
while the $\eta$ parameter can be made small by carefully tuning the parameter $A$.
Any higher order corrections to the K\"ahler potential do not contribute to the leading contributions in the expansion near $\rho=0$ for $\eta$ and $\epsilon$.
Such corrections can therefore be used to alter the potential near its minimum, at some point $X \neq 0$ without influencing the slow-roll parameters.

\subsubsection*{A comment on the $\eta$-problem in Supergravity}

A few words are now in order concerning the $\eta$-problem~\cite{Copeland:1994vg}.
The $\eta$ problem in $\mathcal N = 1$ supergravity is often stated as follows (see for example~\cite{Baumann:2014nda}):
If, for instance, a theory with a single chiral multiplet with scalar component $\varphi$ is taken,
then the K\"ahler potential can be expanded around a reference location $\varphi = 0$ as 
$\mathcal K = \mathcal  K (0) + \mathcal  K_{\varphi \bar \varphi} (0) \varphi \bar \varphi + \dots $ . The Lagrangian becomes
\begin{align}
 \mathcal L = - \partial_\mu \phi \partial^\mu \bar \phi  - \mathcal V(0) \left(1 +  \kappa^2 \phi \bar \phi + \cdots \right) ,
\end{align}
where $\phi$ is the canonically normalised field $\phi \bar \phi = \mathcal  K_{\varphi \bar \varphi} (0) \varphi \bar \varphi$, 
and the ellipses stand for extra terms in the expansion coming from $\mathcal K$ and $W$. 
Following this argument, the mass $m_\phi$ turns out to be proportional to the Hubble scale
\begin{align}
 m_\phi^2 = \kappa^2 \mathcal V(0) + \dots = 3 H^2 + \dots ,
\end{align}
and therefore 
\begin{align}
 \eta  = \frac{m_\phi^2}{3 H^2} = 1 + \dots .
\end{align}
Or otherwise stated, this leading contribution of order 1 to the $\eta$-parameter has its origin from the fact that the F-term contribution to the scalar potential
contains an exponential factor $e^{\mathcal K}$: $\mathcal V = e^{X \bar X + \dots }\left[ \dots \right]$ resulting in its second derivative $\mathcal V_{X\bar X } = V [ 1 + \dots ]$.

However, in our model the factor '1' drops out for the particular choice $b=1$ in the superpotential\footnote{
Note that in hybrid inflation models the $\eta$-problem is also evaded by a somewhat similar way, 
but these models generally include several scalar fields (and superfields) besides the inflaton (see e.g.~\cite{Dvali:1994ms}).}, 
resulting in an inflaton mass $m^2_\rho$ which is determined by the next term $A (X\bar X)^2$ in the expansion of the K\"ahler potential,
\begin{align}
  m_{\chi}^2 &= \left( -4 A  + x^2 \right) \kappa^{-2} f^2 + \mathcal O(\rho^2), \notag  \\
  H^2 &= \frac{ \kappa^{-2} f^2}{6} (2 + x^2) + \mathcal O(\rho^2) . \label{rhomass}
\end{align}
As a result, there are two ways to evade the $\eta$-problem:
\begin{itemize}
\item
First, one can obtain a small $\eta$ by having a small $q \ll f$, while $A$ should be of order~$\mathcal O (10^{-1})$.
In this case, the r\^ ole of the gauge symmetry is merely to constrain the form of the K\"ahler potential and the superpotential, 
and to provide a Higgs mechanism that eliminates the extra scalar (phase) degree of freedom.
\item
Alternatively there could be a cancellation between $q^2$ and $4Af^2$. 
\end{itemize}
Since $A$ is the second term in the expansion of the K\"ahler potential eq.\,(\ref{case1_defs}),
it is natural to be of order~$\mathcal O (10^{-1})$ and therefore providing a solution to the $\eta$-problem.
This will be demonstrated in an example in Section \ref{sec:case1_xi}. 

Note that the mass of the inflaton given in eqs.\,(\ref{rhomass}) is only valid during inflation at small $\rho$. 
The mass of the inflaton at its VEV (Vacuum Expectation Value) will be affected by additional corrections that are needed to obtain in particular a vanishing value for the scalar potential at its minimum, which will be discussed in the following sections.  

\subsubsection*{The upper bound on the tensor-to-scalar ratio}

Before moving on to the next section, let us focus on the approximation at $\rho \ll 1$ where the perturbative expansion of the slow-roll parameters in eqs.\,(\ref{case1_slowroll}) is valid, and assume that the horizon exit occurs at the field value $\rho_*$ very close to the maximum $\rho = 0$.  In this approximation, eqs.\,(\ref{case1_slowroll}) become
\begin{equation}
\label{SRperturbative}
\epsilon (\rho)\approx \epsilon^{\rm pert}(\rho)=|\eta_*|^2\rho^2 , \quad \eta(\rho) \approx \eta_*,
\end{equation}
where the asterisk refers to the value of parameters evaluated at the horizon exit.

To discuss the upper bound on the tensor-to-scalar ratio, it is convenient to divide the region $[\rho=0,\rho_{\rm end}]$ into two regions: one is $[0,\rho_{\rm p}]$, where the approximation \eqref{SRperturbative} is valid, and the other is the rest $[\rho_{\rm p}, \rho_{\rm end}]$. Here $\rho_{\rm end}$ means the inflation end.
Note that $\rho_{\rm p}<\rho_{\rm end}$ because the approximation \eqref{SRperturbative} breaks down before the end of inflation where $\epsilon(\rho_{\rm end}) = 1$ or $|\eta(\rho_{\rm end})| = 1$.
In terms of this division, the number of e-folds from the horizon exit to the end of inflation can be approximated by
\begin{equation}
\label{NCMB-approx}
N_{\rm CMB} \simeq N^{\rm pert}(\rho_*,\rho_{\rm p}) + \kappa \int_{\chi_{\rm p}}^{\chi_{\rm end}} \frac{d \chi}{\sqrt{2\epsilon(\chi)}},
\end{equation}
where we introduced
\begin{equation}
N^{\rm pert}(\rho_1,\rho_2) = \kappa \int_{\chi_1}^{\chi_2} \frac{d \chi}{\sqrt{2\epsilon^{\rm pert}(\chi)}}=\frac{1}{|\eta_*|} \ln \left( \frac{\rho_2}{\rho_1} \right).
\end{equation}
Here $\chi$ is the canonically normalised field defined by eq.\,(\ref{case1_norm}).
Let us next focus on the region $[\rho_{\rm p},\rho_{\rm end}]$.
It is natural to expect the following inequality
\begin{align}
\label{inequality}
\kappa \int_{\chi_{\rm p}}^{\chi_{\rm end}} \frac{d \chi}{\sqrt{2\epsilon(\chi)}} \lesssim \kappa \int_{\chi_{\rm p}}^{\chi_{\rm end}} \frac{d \chi}{\sqrt{2\epsilon^{\rm pert}(\chi)}}.
\end{align}
This is based on the following observation.
The right hand side describes a hypothetical situation, as if the slow-roll condition were valid throughout the inflation until its end. But since in the actual inflation the slow-roll condition breaks down in the region $[\rho_{\rm p},\rho_{\rm end}]$, the actual number of e-folds in this region will be smaller than that in the hypothetical situation.
Adding $N^{\rm pert}(\rho_*,\rho_{\rm p})$ to the both hand sides of \eqref{inequality} and using \eqref{NCMB-approx}, we find
\begin{equation}
N_{\rm CMB} \lesssim \frac{1}{|\eta_*|} \ln \left( \frac{\rho_{\rm end}}{\rho_{*}} \right).
\end{equation}
Using \eqref{SRperturbative} and the definition of the tensor-to-scalar ratio $r = 16 \epsilon_*$, we obtain the upper bound:
\begin{equation}
r \lesssim 16 \left(|\eta_*| \rho_{\rm end} e^{-|\eta_*| N_{\rm CMB} }\right)^2.
\end{equation}
To satisfy CMB data, let us choose $\eta = -0.02$ and $N_{\rm CMB} \approx 50$.  Assuming $\rho_{\rm end} \lesssim 1/2$, we obtain the upper bound $r \lesssim 10^{-4}$. 
Note that this is a little bit lower than the Lyth bound~\cite{Lyth} for small field inflation,  $r \lesssim 10^{-3}$. 
From the upper bound on $r$, we can also find the upper bound on the Hubble parameter as follows. 
In general, the power spectrum amplitude $A_s$ is related to the Hubble parameter at horizon exit $H_*$ by
\begin{align}
 A_s &= \frac{2\kappa^2 H_*^2}{\pi^2 r}.
\end{align}
Combining this with the upper bound $r \lesssim 10^{-4}$ and the value $A_s = 2.2 \times 10^{-9}$ by CMB data, we find the upper bound on the Hubble parameter $H_* \lesssim 10^9$ TeV.

In Section \ref{sec:no_TeV}, we will also find the lower bound $r \gtrsim 10^{-9}$ (equivalently $H_* \gtrsim 10^7$ TeV), based on an model-independent argument. This bound can be lowered at the cost of naturalness between parameters in the potential.


\subsection{de Sitter vacua}

In the previous section we showed that for these models the slow-roll parameters can be small if 
\begin{align}
\frac{-4Af^2 + q^2}{2 f^2 + q^2} \ll 1 \label{case1_slowrollsmall} 
\end{align}
  in eqs.\,(\ref{case1_slowroll}).
We showed above that it is indeed easy to satisfy this condition, providing a potential solution to the $\eta$-problem.
In this set-up however, we are ignorant of what happens after the end of inflation. 
In the following, we will extend the above analysis to include the minimum of the potential, 
and we require after inflation the scalar field $\rho$ to roll into a `nearby' minimum of the potential with a tunably small but positive cosmological constant.
We show below that, for both cases $q = 0$ and $q \neq 0$, that the slow-roll conditions do not allow for the existence of such a minimum.

In other words, while it is easy to safisfy the slow-roll conditions at the maximum of the potential, 
additional corrections are needed to ensure a vanishing (or tunably small and positive) cosmological constant.
These corrections terms modify the potential near the minimum, while the leading contributions to the slow-roll parameters, given by eqs.\,(\ref{case1_slowroll}), 
only depend on the first two terms in the expansion of the K\"ahler potential $\mathcal K = X \bar X + A (X \bar X)^2 + \dots $.
Moreover, as is often the case, having a vanishing cosmological constant comes at the cost of fine-tuning parameters in the model.
An example of such a correction is proposed in Section \ref{sec:case1_xi} and compared with the most recent CMB results.

Let us here explain more precisely our definition for a minimum `nearby' the maximum. 
In a phenomenologically realistic model, inflation should end before the minimum of the scalar potential.  We mentioned in previous sections that the perturbative expansion of the slow-roll parameters in eq.\,(\ref{case1_slowroll}) is not valid at the end of inflation. So, it must not be valid at the minimum either. We should then define the existence of a nearby minimum around the maximum of the potential in the `weaker' sense in which the minimum is in a region where the perturbative expansion of the K\"ahler potential is valid but not of the slow-roll parameters.

\subsubsection*{The need for additional corrections for $ q =0$}

In this section we assume $q=0$ (the case $q\neq 0$ is treated separately below), 
and we show that a model defined by eqs.\,(\ref{case1_defs}) does not allow for a tunably small cosmological constant at a nearby minimum when $|\eta| \ll 1$ at the maximum is required.

\begin{figure}[h!]
    \centering
            \includegraphics[width=0.6\textwidth]{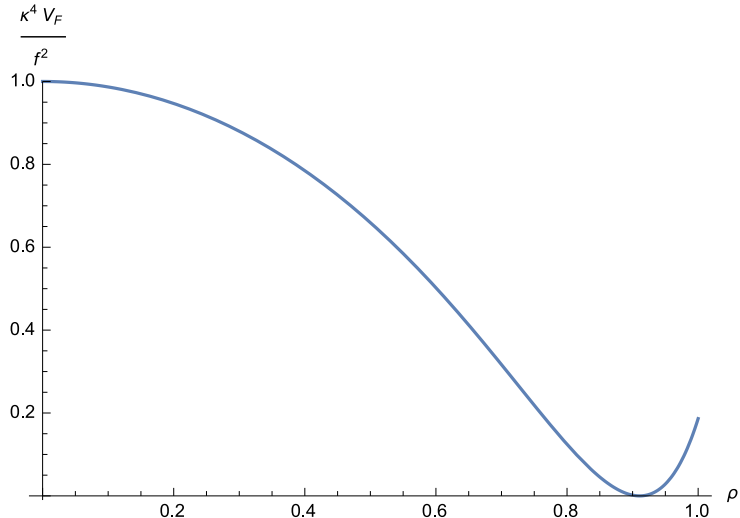}
     \caption{A plot of the scalar potential given by eq.\,(\ref{scalarpot_VF})} satisfying eqs.\,(\ref{VF_sols}).
\label{Fig_VF}
\end{figure}

The scalar potential $\mathcal V = \mathcal V_F$ given by eq.\,(\ref{case1_scalarpot}) with $q=0$,  and is repeated here for convenience
\begin{align}
  \mathcal V = 
\kappa^{-4} f^2 e^{\rho^2 + A \rho^4} 
 \left( - 3 \rho^2 + \frac{\left( 1 + \rho^2 + 2 A \rho^4\right)^2}{1 + 4 A \rho^2} \right) . \label{scalarpot_VF}
\end{align}
The solutions to $\mathcal V (\rho_0) = 0$ and $\mathcal V ' (\rho_0) = 0$ give
\begin{align}
 \rho_0 = \pm 0.91082 , \quad
 A = 0.330858. \label{VF_sols}
\end{align}
A plot of this potential is shown in Figure~\ref{Fig_VF}.
The inflaton starts near the local maximum at $\rho = 0$ and rolls towards the minimum at $\rho =  0.91082$.
Near the local maximum of the potential at $\rho = 0$, $\eta$ is given by eq.\,(\ref{case1_slowroll}) while the parameter $A$ is already fixed by requiring a Minkowski vacuum
in eq.\,(\ref{VF_sols}). As a result, we have
\begin{align}
 \eta &=  - 4 A \approx -1.32 .
\end{align}
We conclude that the slow-roll condition $|\eta| \ll 1 $ is not consistent with the existence of a tunably small cosmological constant at a nearby minimum for $q=0$.


\subsubsection*{The need for additional corrections for $q \neq 0 $} \label{appendix:qneq0}

The scalar potential is given by eq.\,(\ref{case1_scalarpot}), and is repeated here for convenience
\begin{align}
 \mathcal V = 
 \kappa^{-4} f^2 e^{\rho^2 + A \rho^4} 
 \left( - 3 \rho^2 + \frac{\left( 1 + \rho^2 + 2 A \rho^4\right)^2}{1 + 4 A \rho^2} \right)
 + \frac{q^2}{2} \left( 1 + \rho^2 + 2 A \rho^4 \right)^2 . \label{case1_scalarpot_repeated}
\end{align}
The equation $\mathcal V' (\rho) = 0$ is solved by 
\begin{align}
    - \frac{q^2 \left(1 + 4 A \rho^2 \right)^3 }{f^2 e^{\rho^2 + A \rho^4} }  =   
    -4A + (1-6A) \rho^2 + (8A - 16 A^2) \rho^4 + 20 A^2 \rho^6 + 16 A^3 \rho^8. 
\label{appendix:sol1}
   \end{align}
The other solutions to $\mathcal V'(\rho) = 0$ are $\rho = 0$, which is assumed to be a maximum near the start of inflation,
and $1 + \rho^2 + 2 A \rho^4 = 0$ which corresponds to an AdS (Anti-de Sitter) minimum (if $A < 0$). We therefore focus on the solution eq.\,(\ref{appendix:sol1}) with $A > 0$ .

The condition for a vanishing value of the scalar potential at the minimum $\mathcal V (\rho) = 0$ can be combined with eq.\,(\ref{appendix:sol1}) to yield
\begin{align}
 0 = \left( 1 + \rho^2 + 2 A \rho^2 \right)^2 
 - \frac{2 \left( 1 + 4 A \rho^2 \right)^2 \left(  1 - \rho^2 + \rho^4 (1- 8 A) + 4 A \rho^6 + 4 A^2 \rho^8 \right) }
 {-4A + (1-6A) \rho^2 + (8A - 16 A^2) \rho^4 + 20 A^2 \rho^6 + 16 A^3 \rho^8  } . \label{rho_0_min}
\end{align}
For $A > 0$, this equation can be solved to give $\langle \rho \rangle = \rho_0$, the value of $\rho$ at the minimum, 
while $\mathcal V (\rho_0) = 0$ is satisfied when the relation between parameters $f$ and $q$ is given by
\begin{align}
 \frac{f^2}{q^2}  &= \mathcal A (A, \rho_0) , \label{fq}
 \end{align}
 where
 \begin{align} 
  \mathcal A (A, \rho)  = - e^{-\rho^2 - A \rho^4}  \left( \frac{\frac{1}{2} \left( 1 + \rho^2 + 2 A \rho^4 \right)^2  }
  {-3 \rho^2 + \left( \frac{ \left( 1 + \rho^2 + 2 A \rho^4 \right)^2 }{ 1 + 4 A \rho^2 } \right) } \right) . 
 \end{align}
 
\begin{figure}[h!]
    \centering
            \includegraphics[width=0.6\textwidth]{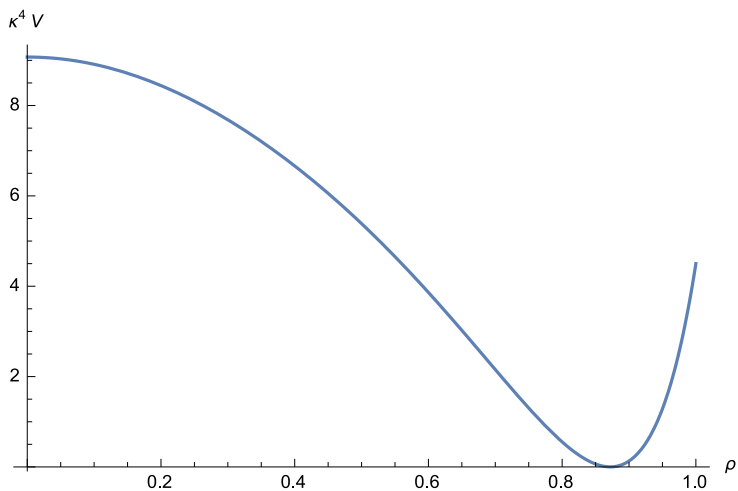}
     \caption{A plot of the scalar potential with $A=1/2$, $q=1$, and $f$ given by eq.\,(\ref{fq}). Notice that the potential indeed has a Minkowski minimum at $\rho = 0.87$. }
\label{Fig_Vq}
\end{figure}

As an example, for $A = \frac{1}{2}$, eq.\,(\ref{rho_0_min}) gives $\rho_0 = 0.872008 $ (the other solutions are not physical).
Taking for example $q = 1$, gives $f \approx 2.928$ by eq.\,(\ref{fq}). 
The result is plotted in Figure~\ref{Fig_Vq}. 

We now show that this is inconsistent with $|\eta| \ll 1$ at the maximum. 
Eq.\,(\ref{case1_slowroll}) can be solved to give
\begin{align}
 \frac{f^2}{q^2} = \frac{2 - \eta}{2 \eta - 8 A}. 
\end{align}
This can be combined with eq.\,(\ref{fq}) to give an expression for $\eta$ in terms of $A$ and $\rho_0$
\begin{align}
 \eta =  2 \frac{1 + A \mathcal A (A , \rho_0) }{ 1 + 2 \mathcal A (A, \rho_0) } .
\end{align}
One can now calculate $\eta$ for every $A>0$ (We do not consider $A<0$ since this would lead to an AdS vacuum).
Numerical analysis however shows that $|\eta| > 1 $ for all $A>0$, and a small $\eta$-parameter is impossible in these models when a vanishing cosmological constant is imposed.

We conclude that additional corrections to the K\"ahler potential are necessary in order to obtain a tunably small cosmological constant consistent with $|\eta| \ll 1$. 
These corrections can be higher order or non-perturbative corrections.


\subsection{Correction terms that allow for a tunable minimum} \label{sec:case1_xi}

We propose corrections to the K\"ahler potential of the form
\begin{align}
 \kappa^2 \mathcal K(X,\bar X) &= X \bar X + \alpha (X \bar X)^2  + F (X \bar X), \label{Kahler_F}
\end{align}
while the superpotential is fixed by the gauge symmetry with $b=1$ (the arguments that excluded $b\neq 1$ in Section \ref{sec:case_1_slowrolls} still apply).
The resulting scalar potential is given by $\mathcal V = \mathcal V_F + \mathcal V_D$, where
\begin{align}
 \mathcal V_F &= \kappa^{-4} f^2 e^{X\bar X + \alpha (X \bar X)^2 + F(X\bar X ) }
 \left[ -3 X \bar X + \frac{ \left| 1 +  X\bar X + 2 \alpha (X \bar X)^2 + X F_X \right|^2 } {1 + 4 \alpha X \bar X + F_{X \bar X} } \right], \notag \\
 \mathcal V_D &=  \kappa^{-4} \frac{q^2}{2} \left| 1 + X \bar X + 2 \alpha (X \bar X)^2 + X F_X \right|^2 ,
 \label{VF_Correction}
\end{align}
where $F_X = \partial_X F$ and $F_{X \bar X} = \partial_X \partial_{\bar X} F$.\footnote{ 
This can in principle  be written in terms of $\rho$ by using $F_X = \frac{1}{2} F_\rho e^{-i\theta}   $ and $F_{X\bar X} = \frac{1 }{4}F_{\rho \rho} + \frac{1 }{4\rho} F_{\rho}$. }

\begin{figure}[h!]
    \centering            \includegraphics[width=0.6\textwidth]{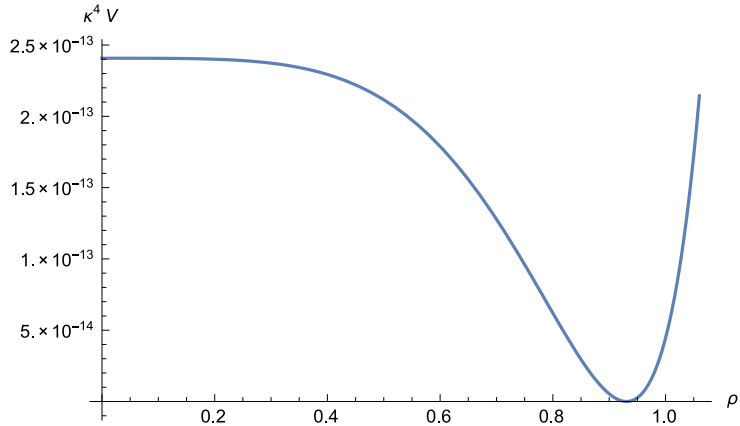}
     \caption{A plot of the scalar potential given by eq.\,(\ref{VF_Correction}) for the parameters in eqs.\,(\ref{case1_pars}).}\label{Fig_Vxi}
\end{figure}

For example, if we choose
\begin{align}
 F(X \bar X) = \xi X \bar X e^{B X \bar X}, \label{F_example}
\end{align}
the K\"ahler potential can be expanded near $X=0$ as
\begin{align}
 \kappa^2 \mathcal K (X \bar X) &= (1 + \xi) X \bar X + (\alpha + \xi B) (X \bar X)^2 + \dots \notag \\
 &= X' \bar X' + \frac{\alpha + \xi B}{ (1 + \xi)^2} (X' \bar X')^2 + \dots ,
\end{align}
where we made a field redefinition $X' = \sqrt{ 1 + \xi} X$ to bring the K\"ahler potential into the form of the expansion eqs.\,(\ref{case1_defs}).\footnote{
Note that we also have $W = \frac{1}{1 + \xi} f X'$, although this is not important for the discussion below.} 
It is important to emphasise that the higher order corrections to the K\"ahler potential do not contribute to the leading order of slow-roll parameters.
One can therefore apply  eqs.\,(\ref{case1_slowroll}) with $A = \frac{\alpha + \xi B}{ ( 1 + \xi )^2 }$ 
 to find
\begin{align}
 \epsilon &= 4 \left( \frac{-4 A + x^2}{2 + x^2} \right)^2  \rho'^{ \ 2} =  4 \left( \frac{ \mathcal B + x^2}{2 + x^2} \right)^2 (1+\xi) \rho^2  ,  \notag \\
 \eta &=  -4 A =  \mathcal B, 
\end{align}
where we used that $\rho'^{ \ 2} = (1+\xi) \rho^2$, and
\begin{align}
 \mathcal B =  -4 \frac{ \alpha   + B  \xi   }{  (1 + \xi)^2 } .
\end{align}
The scalar potential for the parameters 
\begin{align} \alpha=0.41193, ~~ \xi= 0.26790, ~~ B = - 1.51910, ~~ f = 5.52 \times 10^{-7}, ~~ x=0.0526, \label{case1_pars} \end{align}
is plotted in Figure~\ref{Fig_Vxi}. 
\begin{figure}[h!]
    \centering            \includegraphics[width=0.48\textwidth]{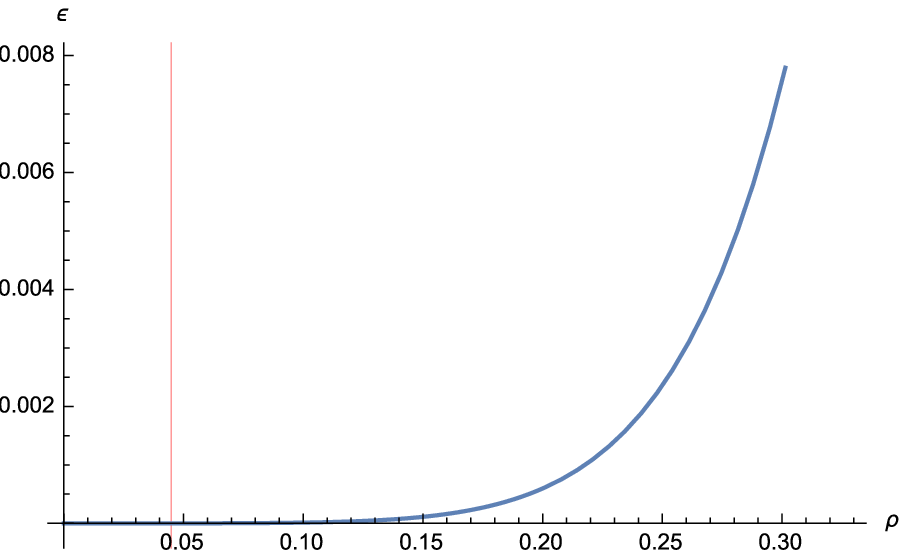}  \includegraphics[width=0.48\textwidth]{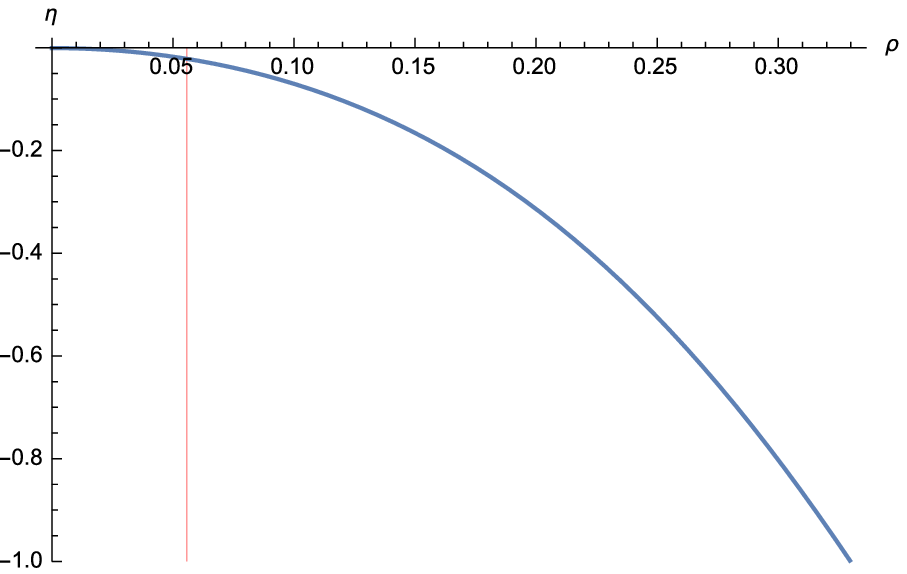}
         \caption{A plot of $\epsilon$ and $\eta$ versus $\rho$ for the scalar potential given by eq.\,(\ref{VF_Correction}) with the parameters in eqs.\,(\ref{case1_pars}). The vertical (red) line indicates $\rho_*$.}\label{Fig_Eta}
\end{figure}
By using $\rho_* = 0.04493$ and $\rho_{\rm end} = 0.32844 $ one obtains the results in Table \ref{prediction_q0}. The slow-roll parameters $\epsilon$ and $\eta$ during the inflation are shown in Figure~\ref{Fig_Eta}. By using eq.~(\ref{slowroll_pars2}), the value of the slow-roll parameters at the horizon exit are 
\begin{equation}
\epsilon(\rho_{*}) \simeq 4.58 \times 10^{-7} \text{ and } \eta(\rho_{*}) \simeq -2.25 \times 10^{-2}.
\end{equation}
\begin{table}
  \centering 
  \begin{tabular}{|c|c|c|c|}
\hline
$N$ & $n_s$ & $r$ & $A_s$  \\
\hline
 57.98 & 0.9549 & $7.3241\times 10^{-6} $ & $2.22  \times 10^{-9} $  \\
\hline
\end{tabular}
  \caption{The theoretical predictions for the parameters given in Figure~\ref{Fig_Vxi}. }
  \label{prediction_q0}
\end{table}

\begin{figure*}
\begin{center}
  \includegraphics[width=0.7\linewidth]{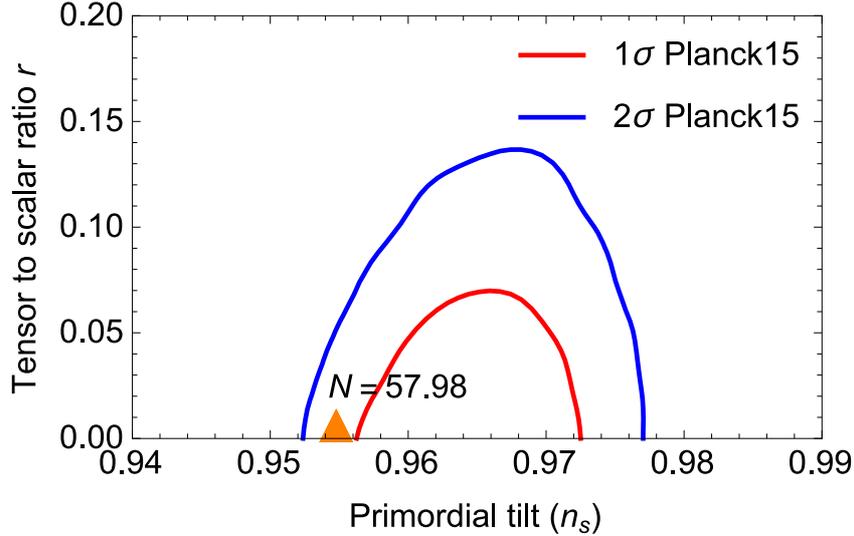}
 \caption{A plot of the predictions shown in Table~\ref{prediction_q0} for a scalar potential in eq.\,(\ref{VF_Correction}) with correction term eq.\,(\ref{F_example}), 
 in the $n_s-r$ Planck '15 results for TT,TE,EE,+lowP and assuming $\Lambda\text{CDM} + r$.}
 \label{n_r_plot_case1}
\end{center}
\end{figure*}

It is important to emphasise that neither the choice of function $F$ in eq.\,(\ref{F_example}), nor the choice for the particular values of the parameters is unique.
Different choices of $F$, or even different choices for the parameters $\alpha$, $\xi$ and $B$ can lead to similar CMB results as the ones presented in Table~\ref{prediction_q0}.
Figure~\ref{n_r_plot_case1} shows that our predictions for $n_s$ and $r$ are within $2 \sigma$ C.L. of Planck '15 contours with the number of e-folds $N=57.98$.

On the other hand, the gravitino mass is given by $m_{3/2} = 5.96 \times 10^7 \text{ TeV}$ and the Hubble scale is $6.80 \times 10^8 \text{ TeV}$. 
No other parameter set consistent with the CMB data were found with a Hubble scale below $10^8 \text{ TeV}$.
Moreover, the tensor-to-scalar ratio satisfies $r> 10^{-6}$. 
We believe that the inability to find models with gravitino mass of order $10 \text { TeV}$ is independent of the choice of $F$. 
We will put stringent conditions on the existence of models with a TeV scale Hubble parameter in Section \ref{sec:no_TeV}.
This will be done in a way not leaning on any ingredients of supergravity, and these results are therefore more general than the scope of this paper. 
However, in supergravity the gravitino mass is usually of the same order as the Hubble scale, which heavily constraints the supersymmetry breaking scale.

In particular, in Section \ref{sec:no_TeV} we show that a necessary, but not sufficient condition to obtain order 10 TeV scale inflation is given in eq.\,(\ref{V_Assumption}),
repeated here for convenience of the reader
\begin{align}
\frac{1}{\kappa^3}  \left| \frac{V_* '''}{V_*} \right| > 10^6 , \label{V_asumption_repeated}
\end{align}
where the derivatives are with respect to the canonically normalised field $\chi$ satisfying eq.\,(\ref{case1_norm}).
This can for example be realised by including an extra term in the K\"ahler potential proportional to $Z (X \bar X)^3$.
The above constraint then requires $Z \gg 1$, 
which violates our assumption that the K\"ahler potential can be expanded around the maximum of the potential at the symmetric point $X = 0$.
As a result, the expansions for $\epsilon$ and $\eta$ in eqs.\,(\ref{case1_slowroll}) require correction terms of higher order in $\rho$,
and the above analysis is not valid. 
Moreover, the above condition~(\ref{V_asumption_repeated}) forces $Z$ to be several orders of magnitude larger than the other parameters, 
which raises questions about naturalness of such a model. 

The authors confirm that the inclusion of the term $Z (X \bar X)^3$ indeed leads to potentials that allow a Hubble scale of order $10$~TeV for large $Z$.
The discussion of such models however is postponed for future work, since the inclusion of a very large parameter is not natural, 
and such a term would reintroduce higher order terms in $\rho$ in the above analysis.

A final comment is in order: 
In contrast with Case 2 in Section \ref{sec:case_2}, 
where the D-term and the F-term contributions to the scalar potential are of the same order
such that a vanishing cosmological constant can be found by a careful cancellation between the F-term and the D-term,
the model above allows for a vanishing cosmological constant even when $q^2 \ll f^2$. 
A vanishing cosmological constant can be obtained by carefully tuning the parameters governing the F-term contribution to the scalar potential.
However, we emphasise that even in this scenario the gauge symmetry still plays an important r\^ ole: 
It constrains the form of the K\"ahler potential and the superpotential, while the $U(1)$ gauge boson `eats' the phase of $X$ resulting in an inflation scenario with a single scalar degree of freedom.\footnote{
For other approaches involving a single scalar degree of freedom see~\cite{FKLP}, or~\cite{SingleScalar} for more recent work.}

\subsection{Anomaly cancellation} \label{sec:Anomaly}

In this section we show that the coefficient $\beta$ in eq.\,(\ref{case1:gaugekin}) is indeed small and can be neglected.
First, as was already noted in eq.~(\ref{Wtransform}), the superpotential is not gauge invariant under the $U(1)_R$ gauge symmetry, and therefore is an R-symmetry.
A consequence of $U(1)_R$ is that scalars and fermions within the same multiplet carry different charges under the symmetry.
In particular,
the total $U(1)_R$-charges of the chiral fermion $\chi$, the $U(1)_R$-gaugino and the gravitino are given by
\begin{align}
 R_\chi = -q + \frac{\xi_R}{2} = - \frac{q}{2}, \quad
 R_\lambda =  R_{3/2} = - \frac{q}{2}. 
\end{align}
This results in a contribution to the cubic $U(1)_R^3$ anomaly proportional to $\mathcal C_R = \Tr [R^3] = R_\chi^3 + R_\lambda^3 + 3 R_{3/2}^2$ \footnote{
Note that the contribution of the gravitino is three times that of a gaugino~\cite{Nielsen:1978ex}.}, given by~\cite{Antoniadis:2014iea}
\begin{align}
 \delta \mathcal L_{1-loop} = - \frac{w }{ 32 \pi^2 } \frac{\mathcal C_R}{3} \epsilon^{\mu \nu \rho \sigma} F_{\mu \nu} F_{\rho \sigma} .
\end{align}
This can however be canceled by a Green-Schwarz mechanism, since the field-dependent term in gauge kinetic function eq.\,(\ref{case1:gaugekin}) results in a contribution to the Lagrangian of the form
\begin{align}
 \mathcal L_{GS} = \frac{1}{8} \text{Im} f (X) \epsilon^{\mu \nu \rho \sigma} F_{\mu \nu} F_{\rho \sigma}
 = \frac{1}{8} \beta_R \text{Im} ( \ln X )  \epsilon^{\mu \nu \rho \sigma} F_{\mu \nu} F_{\rho \sigma} .
\end{align}
Under a gauge transformation, this results in a contribution
\begin{align}
\delta \mathcal L_{GS} = - \frac{w}{8} \beta_R q    \epsilon^{\mu \nu \rho \sigma} F_{\mu \nu} F_{\rho \sigma} .
\end{align}
Anomaly cancelation requires $\delta \mathcal L_{1-loop} + \delta \mathcal L_{GS} = 0$,
\begin{align}
 \beta_R = - \frac{\mathcal C_R}{12 \pi^2 q} .
\end{align}
Contributions from MSSM fermions and mixed anomalies can be treated similarly.

If we instead assume that no extra matter fields are present, we have $\mathcal C_R = - 5 \left( \frac{q}{2} \right)^3$, and $\beta_R = \frac{5}{96 \pi^2} q^2$. 
It follows that $\beta_R$ is very small if $q$ is small.
For example, in the previous section we had $ q = 2.91 \times 10^{-8} $ and $ \rho_{*} = 0.04493$, which gives 
$ \left| \beta_R \ln \rho_{*} \right| \approx 1.8 \times 10^{-17} \ll 1, $  
which justifies the approximation 
\begin{align}
f(X) = 1 + \beta_R \ln X \approx 1.                                   
 \end{align}


\section{Case 2: Inflation away from the symmetric point} \label{sec:case_2}

In this section we consider the case where inflation starts at a point far away from the symmetric point $X=0$.
This allows us to make a field redefinition eq.\,(\ref{field_redef}), and work with a chiral superfield $S$ (with scalar component $s$) invariant under a shift symmetry. 
In this case, the most general superpotential is given by $W= \kappa^{-3} a e^{bs}$, while the K\"ahler potential should be a function of $s + \bar s$. 
The gauge kinetic function is at most linear in $s$ i.e. $f(s) = \gamma + \beta s$.  

Since we are focused on small-field inflation, it is natural to assume that the inflaton starts near a (local) maximum of the potential\footnote{
For small-field inflation, the Lyth bound prevents the possibility of a monotonically increasing potential since this would require $\eta > 0$.
Moreover, inflation starting near an inflection point at a field value greater than the minimum of the potential would also require $\eta > 0$.
On the other hand, inflation starting near an inflection point at a field value smaller than its VEV is in principle allowed. 
However, we could not find any particular working example and therefore focus on inflation starting near a maximum.	}.
However, in contrast with Case 1, 
the maximum of the potential does not correspond to a point of restored R-symmetry and there is no particular reason why the inflation started precisely near this point.
This poses a fine-tuning problem that is present in many small-field models of inflation, and we shall not address this further.
We will show that, 
as in the case 1, the minimum of the inflation potential lies near the local maximum in the weaker sense.


\subsection{Behaviour near the maximum of the potential} 
Let us start by considering a perturbation around the local maximum of the scalar potential $\phi_0$, or equivalently making the change of variable $s = \delta s + \phi_0/2$.  We expand the K\"ahler potential around the maximum up to the 4th order of the classical fluctuation $\delta s$ (and $\delta \bar{s}$)
\begin{equation}
\mathcal{K}(\delta s,\delta\bar{s}) = \kappa^{-2} (\tilde{A} + \tilde{B}(\delta s+\delta\bar{s})+\tilde{C}(\delta s+\delta\bar{s})^2+\tilde{D}(\delta s+\delta\bar{s})^3+\tilde{E}(\delta s+\delta\bar{s})^4 ).
\label{K_expansion_case2}
\end{equation}
We will show below that higher-order terms in the expansion of the K\"ahler potential do not contribute to the ``leading order" terms in slow-roll parameters. 
Indeed, it was already emphasised in the previous section that only terms up to $(X\bar X)^2$ are needed in the K\"ahler potential.
We can also express the superpotential and gauge kinetic function as
\begin{equation}
W(\delta s) = \kappa^{-3} a e^{b(\delta s+\phi_0/2)},
\end{equation}
\begin{equation}
f(\delta s) = \gamma + \beta(\delta s + \phi_0/2).
\end{equation}
By a K\"ahler transformation to absorb $\tilde{A}$ followed by the redefinitions
\begin{align}
 \gamma' = \gamma + \beta\frac{\phi_0}{2}, \quad
 a' = a e^{\tilde{A}/2} e^{b\phi_0 /2}, \quad
 b' = b + \tilde{B},
 \label{abprime}
\end{align}
the K\"ahler potential and superpotential can be simplified into
\begin{align}
\mathcal{K}(\delta s,\delta \bar{s}) &= \kappa^{-2} (\tilde{C}(\delta s+\delta\bar{s})^2+\tilde{D}(\delta s+\delta \bar{s})^3+\tilde{E}(\delta s+\delta \bar{s})^4 ), \notag \\
W(\delta s) &= \kappa^{-3} a e^{b\delta s}, \notag \\
f(\delta s) &= \gamma + \beta \delta s,
\end{align}
where we dropped the primes on $a,b,\gamma$. The D-term contribution to the scalar potential is given by
\begin{equation}
V_D = c^2\kappa^{-4} \frac{\left( b+2\tilde{C}(\delta s+\delta\bar{s})+3\tilde{D}(\delta s+\delta \bar{s})^2+4\tilde{E}(\delta s+\delta\bar{s})^3 \right)^2}{2\gamma + \beta (\delta s+\delta\bar{s})},
\end{equation}
and the F-term contribution is
\begin{eqnarray}
V_F &=& \frac{a^2}{2\kappa^{4}}e^{\tilde{C}(\delta s+\delta \bar{s})^2+\tilde{D}(\delta s+\delta \bar{s})^3+\tilde{E}(\delta s+\delta\bar{s})^4+b(\delta s+\delta\bar{s})} \nonumber\\
&& \times \left(
\frac{(b+2 \tilde{C}(\delta s+\delta\bar{s}) +3 \tilde{D}(\delta s+\delta\bar{s})^2+4 \tilde{E} (\delta s+\delta\bar{s})^3)^2}
{\tilde{C}+3 \tilde{D}(\delta s+\delta \bar{s}) +6 \tilde{E} (\delta s+\delta \bar{s})^2}-6\right) .
\end{eqnarray}
Let us define $\phi = s+\bar{s}$ such that $\delta \phi = \delta s + \delta \bar{s}$ represents the (classical) fluctuation around the maximum of the potential at $\phi_0$.  Then the potential can be written in terms of $\delta\phi$ as
\begin{eqnarray}
V &=& \frac{1}{2\kappa^4}\left\{ a^2 e^{\tilde{C} \delta\phi ^2+\tilde{D} \delta\phi ^3+\tilde{E} \delta\phi ^4+b \delta\phi } \left(\frac{(b+2 \tilde{C}\delta\phi+3 \tilde{D}\delta\phi^2+4 \tilde{E}\delta\phi^3)^2}{\tilde{C}+3 \tilde{D}\delta\phi +6 \tilde{E} \delta\phi^2}-6\right)  \right.\nonumber\\
&& + 
\left.\frac{2 c^2 (b+2 \tilde{C}\delta\phi+3 \tilde{D}\delta\phi^2+4 \tilde{E}\delta\phi^3)^2}{2 \gamma+\beta  \delta\phi }\right\} .
\label{case2:Potential}
\end{eqnarray}
We fix $\gamma =1$ and $\beta = 0$ for simplicity.  
The slow-roll parameters $\epsilon$ and $\eta$ are defined in terms of the canonically normalised fluctuation $\delta\chi$, which is defined by
\begin{equation}
d(\delta\chi) = \sqrt{\tilde{C} + 3 \delta\phi (\tilde{D} + 2 \tilde{E} \delta\phi)} d(\delta\phi).
\end{equation}
Since we expand the potential around the maximum of the potential, the slow-roll parameter must satisfy $\epsilon(\delta\phi=0) = 0$.  This gives us the following constraint on the parameters,
\begin{equation}
\tilde{D}=\frac{a^2 b^2 \tilde{C}-2 a^2 \tilde{C}^2+4 c^2 \tilde{C}^3}{3 a^2 b}.
\end{equation}
Using the above constraint, we can write the expansion of the slow-roll parameters near the maximum of the potential as
\begin{align}
 \epsilon &= \frac{\left(a^4 \left(b^4 \tilde{C}-4 b^2 \left(\tilde{C}^2+3 \tilde{E}\right)+12 \tilde{C}^3\right)+12 a^2 c^2 \tilde{C}^3 \left(b^2-4 \tilde{C}\right)+48 c^4 \tilde{C}^5\right)^2}{2 a^4 \tilde{C}^3 \left(a^2 \left(b^2-6 \tilde{C}\right)+b^2 c^2 \tilde{C}\right)^2}\delta\phi ^2 +  \mathcal{O} (\delta\phi^3) , \label{case2:epsilon}
\end{align}
and
\begin{align}
 \eta &= -\frac{a^4 \left(-b^4 \tilde{C}+4 b^2 \left(\tilde{C}^2+3 \tilde{E}\right)-12 \tilde{C}^3\right)-12 a^2 c^2 \tilde{C}^3 \left(b^2-4 \tilde{C}\right)-48 c^4 \tilde{C}^5}{a^2 \tilde{C}^2 \left(a^2 \left(b^2-6 \tilde{C}\right)+b^2 c^2 \tilde{C}\right)} + \mathcal{O}(\delta\phi) .  \label{case2:eta}
\end{align}
Note that $\eta$ parameter can be small and negative by carefully fine-tuning four parameters $a$, $b$, $\tilde{C}$ and $\tilde{E}$.  It is also important to note that the slow-roll expansions are valid only near the maximum of the potential i.e. $\delta\phi \ll 1$.   They may break down during inflation or at the minimum of the potential. These expansions are useful for showing qualitatively that the $\eta$-problem can be avoided.  In order to compare any predictions with the CMB data, one needs to use the full expression for $\epsilon$ and $\eta$.

\subsection{Example}

\begin{figure*}
\begin{center}
  \includegraphics[width=0.7\linewidth]{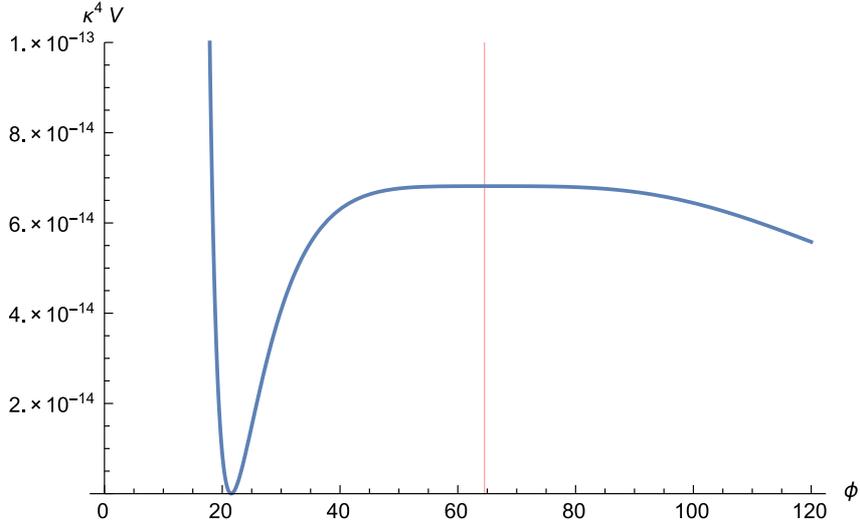}
 \caption{A plot of scalar potential in eq.\,(\ref{case2_potential_p1}), one of the models belonging to case 2 with $\gamma = 1$ and $\beta =0$. We choose  $a = 2.34422 \times10^{-6}$, $b = -0.0234$, $c = 7.10 \times 10^{-6}$, $\xi = 0.3023$, $\alpha = -0.7813$.   The vertical (red) line indicates $\phi_*$.}
 \label{Potential_case2_p1}
\end{center}
\end{figure*}

\begin{figure*}
\begin{center}
  \includegraphics[width=0.7\linewidth]{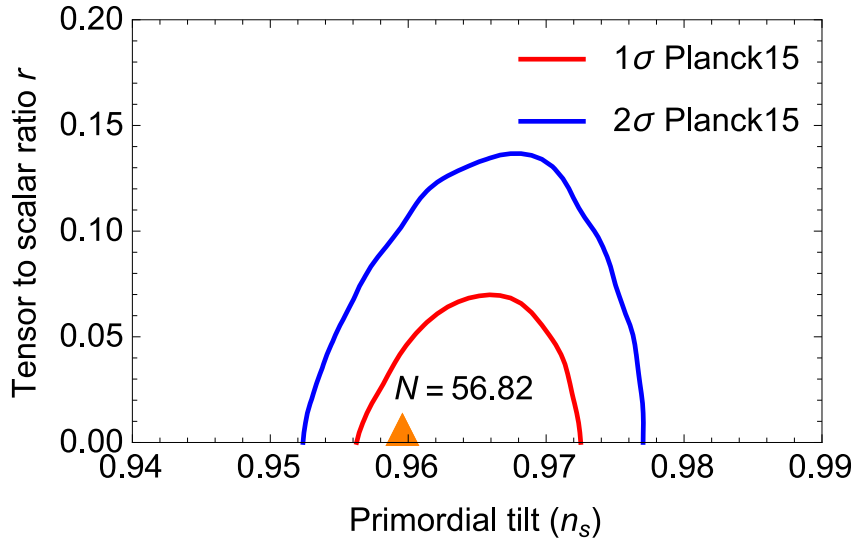}
 \caption{A plot of the predictions for a scalar potential in eq.\,(\ref{case2_potential_p1}) shown in Table \ref{prediction_case2}, in the $n_s$-$r$ Planck '15 results for TT,TE,EE,+lowP and assuming $\Lambda\text{CDM} + r$.}
 \label{n_r_plot_case2_p1}
\end{center}
\end{figure*}

In order to give an example of this class of models, let us consider the case $\gamma =1$, $\beta=0$, with K\"ahler potential \cite{Antoniadis:2016aal}
\begin{equation}
\mathcal{K} = -\kappa^{-2} \ln \left(s+\bar{s}+ \frac{\xi}{b} e^{\alpha b^2 (s+\bar{s})^2}\right).
\label{case2_example}
\end{equation}
This model is obviously invariant under the shift symmetry in eq.~(\ref{shift_symmetry}).  With $\phi =  s+ \bar{s} =\delta\phi + \phi_0$, the potential in terms of fluctuation $\delta\phi$ around the maximum of the potential is given by
\begin{eqnarray}
\mathcal V &=& \frac{\kappa^{-4}b^2 c^2}{2}\left[ \frac{b (\delta\phi + \phi_0) - 1 + \xi e^{\alpha b^2 (\delta\phi + \phi_0)^2}(1-2 \alpha b (\delta\phi + \phi_0) )}{b (\delta\phi + \phi_0) + \xi e^{\alpha b^2 (\delta\phi + \phi_0)^2}}\right]^2\nonumber\\
&{}& -\frac{\kappa^{-4}|a|^2 b e^{b (\delta\phi + \phi_0)}}{  \xi  e^{\alpha  b^2 (\delta\phi + \phi_0)^2}+b(\delta\phi + \phi_0)}\nonumber\\
&&\times \left[\frac{\left( b(\delta\phi + \phi_0)+\xi  e^{\alpha  b^2 (\delta\phi + \phi_0)^2 }(1-2 \alpha b (\delta\phi + \phi_0)) -1\right)^2}{2 \alpha   \xi  e^{\alpha  b^2 (\delta\phi + \phi_0)^2} \left(2 \alpha  b^3 (\delta\phi + \phi_0)^3+\xi  e^{\alpha  b^2 (\delta\phi + \phi_0)^2}-b(\delta\phi + \phi_0)\right)-1}+3\right]. 
\label{case2_potential_p1}
\end{eqnarray}
In this example, the canonically normalised field $\chi$ is defined by
\begin{equation}
\frac{d\chi}{d\phi} = \frac{\kappa^{-1} b }{ \left(\xi  e^{\alpha  b^2 \phi ^2}+b \phi \right)}\sqrt{1-2 \alpha  \xi  e^{\alpha  b^2 \phi ^2} \left(2 \alpha  b^3 \phi ^3+\xi  e^{\alpha  b^2 \phi ^2}-b \phi \right)}.
\end{equation}
For this case, we need contributions from both F-term and D-term in order to obtain Minkowski vacua.  It was shown in \cite{Antoniadis:2016aal,Antoniadis:2014iea} that this model is anomaly-free. 
By choosing, $a = 2.34422 \times10^{-6}$, $b = -0.0234$, $c = 7.10 \times 10^{-6}$, $\xi = 0.3023$, $\alpha = -0.7813$, we obtain an appropriate inflationary potential with a flat plateau around the maximum $\phi_0 \approx 66.37$ illustrated in Figure~\ref{Potential_case2_p1}.
The horizon exit is at $\phi_{*} = 64.5315$ and inflation ends at $\phi_{\rm end} = 50.9915$.  By using eq.~(\ref{slowroll_pars2}), the value of the slow-roll parameters at horizon exit are
\begin{equation}
\epsilon(\phi_{*}) \simeq 1.30 \times 10^{-7} \text{ and } \eta(\phi_{*}) \simeq -2.01 \times 10^{-2}.
\label{slow_roll_example}
\end{equation}
The number of e-folds $N$, the scalar power spectrum amplitude $A_s$, the spectral index of curvature perturbation $n_s$ and the tensor-to-scalar ratio $r$ are calculated and summarised in Table \ref{prediction_case2}, in agreement with Planck '15 data.  Figure~\ref{n_r_plot_case2_p1} shows that our predictions for $n_s$ and $r$ are within $1\sigma$ C.L. of Planck '15 contours with the number of e-folds $N \approx 56.82$. 


During inflation, we can show that a field fluctuation from the maximum of the potential in terms of the canonically normalised field is quite small in Planck units, i.e. $\kappa|\delta \chi_{*}| =\kappa|\chi_{*}-\chi_0| \approx 0.036$ and $\kappa|\delta \chi_{\rm end}| =\kappa|\chi_{\rm end}-\chi_0| \approx 0.351$, but at the minimum of the potential $\phi_{\rm min} \approx 21.50$ the field fluctuation is large, i.e. $\kappa|\delta \chi_{\rm min}| =\kappa|\chi_{\rm min}-\chi_0| \sim 1.81$. However, one can show that the perturbative expansion of the K\"ahler potential is still valid and the minimum is nearby the maximum in the weaker sense. Indeed, using the values of parameters $a$, $b$, $c$, $\xi$ and $\alpha$ given above, we can expand the K\"ahler potential (\ref{case2_example}) in the fluctuation $\delta\phi$ around the maximum of the scalar potential as
\begin{eqnarray}  
\kappa^2\mathcal{K} 
&=& 4.1652 + (1.7256\times10^{-2}) \delta\phi - (1.8499\times10^{-4})\delta\phi^2 +(2.5255\times10^{-6})\delta\phi^3\nonumber\\
&{}& \quad - (3.1815\times10^{-8})\delta\phi^4+(4.0104\times10^{-10})\delta\phi^5-(6.0738\times10^{-12})\delta\phi^6 
\nonumber\\
&{}& \quad
+\mathcal{O}(\delta\phi^7).
\end{eqnarray}
It is easy to prove that this expansion makes sense as a perturbative expansion even for $|\delta\phi_{min}| = |\phi_{min}-\phi_0|\sim 44.9$ (equivalent to $\kappa|\delta\chi_{min}| \sim 1.81$ as given above). Thus, in this example the minimum lies nearby the maximum in the sense defined above. 

Note however that in contrast with the claim in~\cite{Antoniadis:2016aal},
the potential given in eq.~(\ref{case2_potential_p1}) 
is inconsistent with a supersymmetry breaking scale (and Hubble scale) in the multi-TeV range.
It has turned out to be very difficult to find any parameters, consistent with the CMB data, for which the Hubble scale $H_*$ is lower than $10^8$ TeV.
The reason is outlined in Section \ref{sec:no_TeV}: 
Eq.\,(\ref{ns}), defining the amplitude $A_s$, fixes $\epsilon_*$ in terms of $H_*$ (for $A_s$ fixed by the CMB data), resulting in a very small ($\sim 10^{-23}$) value for $\epsilon_*$.
This in turn results in a big number of e-folds occurring near the horizon exit, 
which rapidly exceeds the required amount of e-folds $N_{\rm CMB}$ between the horizon exit and the end of inflation, which should be of order $\sim 40$.
Indeed, the potential in eq.\,(\ref{case2_potential_p1}) does not satisfy the necessary (but not sufficient) condition to obtain order 10 TeV scale inflation $\frac{1}{\kappa^3}\big|\frac{V_*'''}{V_*}\big| \gg 10^6 $ 
in eq.\,(\ref{V_Assumption}).

\begin{table}
  \centering 
  \begin{tabular}{|c|c|c|c|}
\hline
$N$ & $n_s$ & $r$ & $A_s$  \\
\hline
 56.82 & 0.9597 & $2.0747 \times 10^{-6} $ & $2.22  \times 10^{-9} $  \\
\hline
\end{tabular}
  \caption{The theoretical predictions for $\phi_{*} = 64.5316$ and $\phi_{\rm end} = 50.9915 $ and the parameters given in Figure~\ref{Potential_case2_p1}. }
  \label{prediction_case2}
\end{table}


\section{A stringent constraint on TeV Hubble scale} \label{sec:no_TeV}

\mbox{}
In the previous sections, we have seen that it is difficult to realise inflation models with a TeV scale gravitino mass. 
Here we characterise this difficulty in a more general way, and independent of the supergravity framework.
In particular, we show that a necessary but not sufficient condition to have order 10 TeV inflation is that $\frac{1}{\kappa^3}\big|\frac{V_*'''}{V_*}\big| > 10^6 $. 
\mbox{}

The CMB observation covers the inflation between the horizon exit and the end of inflation. We first estimate the number of e-folds during this epoch using the following formula \cite{Liddle:2003as, Martin:2010kz, Gorbunov:2011zzc, Ade:2015lrj}:
\begin{align}
\label{NCMB}
N_{\rm CMB} \simeq \ln \frac{T_0}{q_0}-\left(\frac{4}{y}-1\right)\ln \frac{\sqrt{\kappa^{-1}H_{\rm end}}}{T_{\rm reh}}+\frac{1}{2}\ln (\kappa H_{\rm end})+\ln \frac{H_*}{H_{\rm end}},
\end{align}
where $q_0,T_0$ are the wave number  in the physical spatial coordinate and the CMB temperature at present, respectively. Note that this formula is independent of inflation models. Here we take $T_0/q_0 \sim 10^{28}$ with $q_0 \simeq 0.002 \, {\rm Mpc}^{-1}$ following \cite[pp.325-326]{Gorbunov:2011zzc}. We assume that $H_{\rm end} \simeq H_*$, and that during the reheating era the energy density evolves as $\rho\propto a^{-y}$ with $3\leq y <4$. The asterisk refers to the horizon exit as in the previous sections. 
The reheating temperature $T_{\rm reh}$ is assumed to be of the same order as $H_{\rm end}$. Under the above rough assumptions we find that, for $H_*$ ranging from $1\,{\rm TeV}$ to $10^{10}\,{\rm TeV}$, the number of e-folds $N_{\rm CMB}$ is roughly between 40 and 60.

Let us next discuss the condition that an inflaton potential should yield $N_{\rm CMB}=$ 40 - 60 for a given Hubble parameter $H_*$.
The amplitude $A_s$ and tilt $n_s$ are given by
\begin{align}
 A_s &= \frac{\kappa^4 \mathcal V_*}{24 \pi^2 \epsilon_*} = \frac{\kappa^2 H_*^2}{8 \pi^2 \epsilon_*}, 
 \label{As} \\
 n_s &= 1 + 2 \eta_* -6 \epsilon_*  \label{ns},
\end{align}
where we used that the Hubble scale is given by $H_* = \kappa \sqrt{V_*/3}$. 

\begin{figure}[h!]
    \centering            \includegraphics[width=0.7\textwidth]{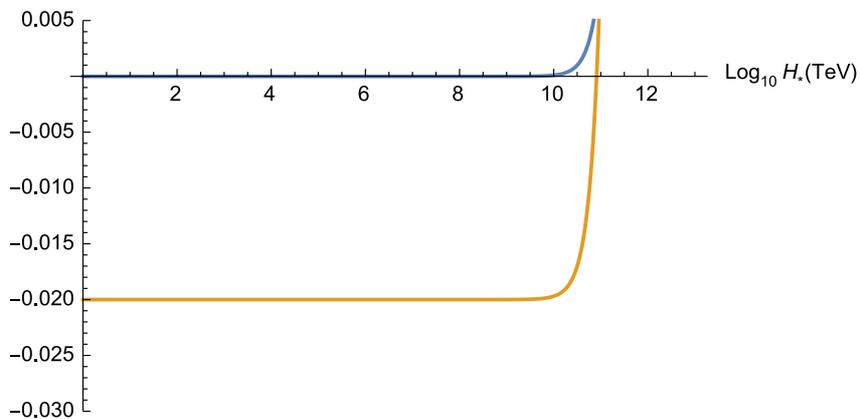}
         \caption{A plot of $\epsilon_*$ and $\eta_*$ versus the Hubble parameter $H_*$ at the horizon exit. The blue line refers to $\epsilon_*$ and the orange one refers to $\eta_*$. Note that $\epsilon_*$ is much smaller than $\eta_*$ for $H_* \lesssim 10^{10} \, {\rm TeV}$, which fits our models given in the previous sections.}
         \label{slowroll-H}
\end{figure}

We choose $A_s = 2.2 \times 10^{-9}$ and $n_s = 0.96$ to satisfy the CMB data.
Putting them into eqs.~\eqref{As} and \eqref{ns} yields the slow-roll parameters $\epsilon_*$ and $\eta_*$ as functions of the Hubble constant $H_*$, given by
\begin{align}
\label{slowrolls}
\epsilon_* \simeq \left( \frac{H_*}{1.0 \times 10^{12} \, {\rm TeV}} \right)^2, \quad
\eta_* \simeq -0.02+\left( \frac{H_*}{5.8 \times 10^{11} \, {\rm TeV}} \right)^2.
\end{align}
The plot of these expressions is given in Figure~\ref{slowroll-H}.
The Lyth bound~\cite{Lyth} for small field inflation, $r \lesssim 10^{-3}$, implies $H_* \lesssim 10^{10} \text{ TeV}$ by eq.~(\ref{slowrolls}).
From Figure~\ref{slowroll-H} it is clear that if the Hubble scale satisfies $H_* \lesssim 10^{10} \, {\rm TeV}$, then the slow-roll parameters at horizon exit satisfy $\epsilon_* \ll |\eta_*|$.\footnote{An upper bound on the tensor-to-scalar ratio $r \lesssim 10^{-4}$ from Section \ref{sec:case_1} implies $H_* \lesssim 10^{9}$ TeV. This is consistent with $H_* \lesssim 10^{10} \, {\rm TeV}$ by the Lyth bound.}  In the following we consider the Hubble scale in the range $1  \, {\rm TeV} \sim 10^{10} \, {\rm TeV}$.\footnote{
Note that the maximum Hubble scale $H_* \sim 10^{10} \, {\rm TeV}$ corresponds to the GUT scale vacuum energy during inflation $V_*^{1/4}\sim M_{\rm GUT}$.}

The inflaton potential decreases monotonically through the end of inflation $\chi_{\rm end}$ towards the global minimum. 
For simplicity and without loss of generality, we assume that the field value of the inflaton decreases monotonically during inflation. We can express the number of e-folds between the horizon exit $\chi_*$ and some field value $\chi$ in terms of the inflaton potential as
\begin{align}
 N(\chi) = \int^{\chi_*}_{\chi} \frac{\kappa}{\sqrt{2 \epsilon(\chi')}} \ d \chi'  .
 \label{e-foldsCMB}
\end{align}
In terms of eq.~(\ref{e-foldsCMB}), the number of e-folds between the horizon exit and the end of inflation is given by $N (\chi_{\rm end}) $.
Any scalar potential should therefore satisfy $N (\chi_{\rm end}) \approx  N_{\rm CMB}$ in order to be a good candidate for slow-roll inflation satisfying the CMB data.

Let us divide the region $[\chi_{\rm end},\chi_*]$ into two parts, region A and B. Region A is in the neighbourhood of the horizon exit $\chi_*$ and the slow-roll parameters in region A are required to satisfy the conditions
\begin{align}
\epsilon(\chi) \ll |\eta(\chi)|, \quad |\eta(\chi)| \approx |\eta(\chi_*)| ,  \label{chi0_assumptions}
\end{align}
and the region B is its complement in $[\chi_{\rm end},\chi_*]$.
Note that the condition \eqref{chi0_assumptions} is natural in the neighbourhood of the horizon exit as long as the Hubble parameter satisfies $H_* \lesssim 10^{10} \, {\rm TeV}$, as was already observed in Figure~\ref{slowroll-H}.
The inflaton first passes through region $A$ and enters region $B$ when the conditions~(\ref{chi0_assumptions}) are violated.
This transition occurs well before the end of inflation where the slow-roll parameters are of order 1.
It is shown in the Appendix that if $\chi$ is in region A, then the number of e-folds $N(\chi)$ can be approximated as 
\begin{align}
N(\chi) \approx \frac{1}{ |\eta_*| } \ln \left[ 1 + \kappa \frac{ \left| \eta_*  \Delta  \chi  \right|}{\sqrt{2\epsilon_*}}  \right] , \label{N_Approx}
\end{align}
where $\Delta \chi = | \chi_* - \chi | $.
Using this expression, we can show that, for any $\chi$ in region A (and therefore satisfy eqs.~(\ref{chi0_assumptions})), 
the following inequality holds:
\begin{align}
N_{\rm CMB} \simeq N(\chi_{\rm end}) 
&\simeq \int_{\chi_{\rm end}}^\chi \frac{\kappa}{\sqrt{2 \epsilon(\chi')}} \ d \chi' + \frac{1}{ |\eta_*| } \ln \left[ 1 + \kappa \frac{ \left| \eta_*  \Delta  \chi  \right|}{\sqrt{2\epsilon_*}}  \right] \nonumber\\
&> \frac{1}{ |\eta_*| } \ln \left[ 1 + \kappa \frac{ \left| \eta_*  \Delta  \chi  \right|}{\sqrt{2\epsilon_*}}  \right]
\equiv N_A(\chi).
\label{criterion}
\end{align}
This inequality can be used to exclude a given potential in the following way: 
If for any particular inflaton potential there exists some $\chi$ in region A for which $N_A(\chi) > N_{\rm CMB}$, then it contradicts the inequality \eqref{criterion}, 
and the inflaton potential can be excluded.
In other words, if the flat region near the start of inflation is too big, 
too many e-folds will occur in region $A$. As a result, the number of e-folds in this region exceeds the required number of e-folds $N_{\rm CMB}$. 
This allows us to put a limit on the scalar potential by investigating the size of region $A$, characterised by $\kappa \Delta \chi$.

 From eq.~(\ref{N_Approx}) we find that,
 for the Hubble scale $H_{*} \sim 10^p \, {\rm TeV}$ with $0 \leq p \leq 10$, 
 the choice $\kappa\Delta\chi \simeq 10^{p-9}$ results in a number of e-folds in region $A$ given by $N_A(\chi)\simeq 80$, 
 which is sufficiently larger than $N_{\rm CMB} \simeq 40$ - $60$.
 It remains to check whether this $\chi$ is still in region $A$ and therefore satisfies eqs.~(\ref{chi0_assumptions}).
    It can be shown\footnote{
Combining $\kappa \Delta \chi = 10^{p-9}$
with the assumptions \eqref{slowrolls} and \eqref{V_Assumption}, we find that
\begin{align}
\Big| \Delta\chi \frac{d\eta}{d \chi} \Big|
= \kappa\Delta\chi \Big| \left(\frac{1}{\kappa^3} \frac{V'''}{V} - \eta \sqrt{2 \epsilon}\right) \Big| \ll |\eta_*|. \nonumber
\end{align}
at the horizon exit.
As a result, the condition $\eta (\chi) \approx \eta (\chi_*)$ satisfied.
Similarly, by Taylor expanding $\sqrt{\epsilon}$ in $\Delta\chi$, we can show that $\epsilon(\chi) \ll |\eta(\chi)|$.
}
 that $\chi$ with $\kappa\Delta\chi \simeq 10^{p-9}$ lies in region A if the inflaton potential satisfies
\begin{align}
\frac{1}{\kappa^3}\bigg|\frac{V_*'''}{V_*}\bigg| \ll 10^{7-p}. \label{V_Assumption}
\end{align} 
In summary, we have shown that if eq.~\eqref{V_Assumption} is satisfied (for $H_* = 10^p \, {\rm TeV}$ with $0 \leq p \leq 10$), 
then there exists $\chi$  in region A which gives $N(\chi)=80$, which exceeds $N_{\rm CMB}$, and therefore contradicts the inequality \eqref{criterion}.
Therefore, the condition in eq.~\eqref{V_Assumption} is a sufficient condition to exclude an inflaton potential.

In other words, the condition 
\begin{align}
\frac{1}{\kappa^3}\bigg|\frac{V_*'''}{V_*} \bigg|  > 10^{7-p}
\end{align}
is a necessary (but not sufficient) condition for any the inflaton potential such that the number of e-folds does not exceed $N_{\rm CMB}$.

A rough estimate on the necessary condition implies that if $H_*  \geq  10^7 \text{ TeV}$, 
a milder condition on the potential $\frac{1}{\kappa^3}\big|\frac{V_*'''}{V_*} \big|  < 1$ should be satisfied.
It should therefore be much easier to construct potentials with a Hubble scale above $10^7 \text{ TeV}$.
Note that this implies a theoretical (rough) lower limit on the tensor-to-scalar ratio, namely $r > 10^{-9}$ (or equivalently, $\epsilon > 10^{-10}$).
On the other hand, if the Hubble scale is low, around the TeV scale, then the necessary condition puts a very stringent constraint on the potential 
which poses questions about the naturalness of the theory since a large ($\sim 10^9$) hierarchy between parameters in the model is necessary.

It is important to emphasise that the above argument holds for all models of (slow-roll) inflation, and is not limited to supergravity models.
In supergravity models however, the Hubble scale $H_*$ is usually of the same order as the gravitino mass $m_{3/2}$.\footnote{
An exception to this occurs in so-called New Inflation models~\cite{New_Inflation}, 
where the Hubble scale is typically much larger than the gravitino mass and a gravitino mass as low as $100 \text{ TeV}$ is allowed~\cite{Harigaya:2013pla}. } 
The above argument therefore puts stringent constraints on the lower limit of the supersymmetry breaking scale. 
It states that a theory with low energy supersymmetry breaking ($m_{3/2}$ of order $\mathcal O (10 \text{ TeV})$)
is very unlikely to simultaneously be a theory for inflation, unless the third derivative of the scalar potential attains very high values near the start of inflation.

\section{Conclusions}

In this work we proposed a class of supergravity models of inflation in which the inflaton is the superpartner of the goldstino, carrying a unit charge under a gauged R-symmetry.
Inflation occurs around the maximum of the scalar potential where R-symmetry is either restored (case 1) or spontaneously broken (case 2), with the inflaton rolling down towards the electroweak minimum of the supersymmetric Standard Model. Analyticity and R-invariance imply a linear superpotential which automatically avoids the so-called eta-problem in supergravity, allowing for slow-roll and small field inflation consistently with an effective field theory description.

Case 2 is a generalisation of a particular model we studied in a previous publication~\cite{Antoniadis:2016aal}, inspired by type I string theory with moduli stabilisation by internal magnetic fluxes and the inflation identified with the string dilaton. Case 1 offers a new possibility, with the R-symmetry restored around the inflationary plateau, which we analysed in detail in this work:
\begin{itemize}
\item
Assuming first that inflation ends very rapidly after the plateau where small fluctuations are valid around the maximum of the scalar potential, we were able to find simple analytic formulae for the slow-roll parameters and the number of e-folds, leading to an upper bound for the tensor-to-scalar ratio $r\simlt 10^{-4}$ and for the inflation scale $H_*\simlt 10^{12}$ GeV.
\item
We then show that additional corrections to the K\"ahler potential are needed in order to describe a minimum of the scalar potential `nearby' the maximum with an infinitesimal tuneable positive vacuum energy. The `nearby' notion is defined in the sense that perturbative expansion around the maximum is valid for the K\"ahler potential but not for the slow-roll parameters. We subsequently study an example of such corrections and work out the experimental predictions for the cosmological observables.
\end{itemize}
We finally derived a stringent and model-independent constraint on TeV scale inflation (or in general low scale). The constraint implies that in order to attain a theory with Hubble scale in the multi-TeV range, one has to tune the form of the potential, implying in general a naturalness problem between its parameters.

\section*{Acknowledgements}
This work is supported in part by Franco-Thai Cooperation Program in Higher Education and Research `PHE SIAM 2016', 
in part by the ``CUniverse'' research promotion project by Chulalongkorn University (grant reference CUAASC).
H.I. would like to thank Toshifumi Noumi for important suggestions.
The authors would like to thank C.~Pongkitivanichkul for stimulating discussions concerning Section~\ref{sec:no_TeV}.

\appendix
\section{Proof of an important identity}

In this Appendix, we show that eq.\,(\ref{N_Approx}) holds for any $\chi$ satisfying eqs.\,(\ref{chi0_assumptions}), repeated here for convenience of the reader
\begin{align}
\epsilon(\chi) \ll |\eta(\chi)|, \quad |\eta(\chi)| \approx |\eta(\chi_*)| . \notag
\end{align}
Note first  that the identity
\begin{align}\label{dsqepdchi}
\frac{ d \sqrt{\epsilon} } {d \chi }  = \frac{1}{\sqrt{2} \kappa} \left( \frac{ V''}{V} - \left( \frac{V'}{V} \right)^2 \right)
 = \frac{\kappa}{\sqrt{2}} \left( \eta - \epsilon \right)
\end{align}
can be approximated between $\chi$ and $\chi_*$ by
\begin{align}
 \frac{1}{\kappa} \frac{ d \sqrt{\epsilon} } {d \chi } &\approx \frac{\eta_*}{\sqrt{2}} . 
 \end{align}
This can be solved analytically by
 \begin{align}
 \sqrt{\epsilon} 
 = \sqrt{\epsilon_*} + \frac{\eta_*}{\sqrt{2}} \kappa ( \chi - \chi_* ).
 \label{epsilon_approx0}
\end{align}
The number of e-folds at $\chi$ then becomes
\begin{align}
N(\chi)=\kappa \int_{\chi}^{\chi_*} \frac{d\chi'}{\sqrt{2 \epsilon(\chi')}}
 \approx \frac{1}{ |\eta_*| } \ln \left[ 1 + \kappa \frac{ \left| \eta_*  \Delta  \chi  \right|}{\sqrt{2\epsilon_*}}  \right].
\end{align}
This completes the proof of eq.\,(\ref{N_Approx}).



\end{document}